\PassOptionsToPackage{table}{xcolor}
\documentclass[10pt, conference]{IEEEtran}
\usepackage{amsmath} 
\usepackage{amsthm} 
\usepackage{amssymb} 
\usepackage{graphicx} 
\usepackage{multicol} 
\usepackage{sparklines} 
\usepackage{subcaption}
\usepackage{cleveref}

\newcommand{\knowledgespark}[5]{
\begin{sparkline}{5}
    \sparkspikewidth 5pt
    \definecolor{sparkspikecolor}{named}{red}
    \sparkspike 0 1.5*#1
    \sparkspike 0.333 1.5*#2
    \sparkspike 0.666 1.5*#3
    \sparkspike 0.999 1.5*#4
    \sparkspike 1.333 1.5*#5
  \end{sparkline}
}

\let\svthefootnote\thefootnote
\newcommand\blankfootnote[1]{%
  \let\thefootnote\relax\footnotetext{#1}%
  \let\thefootnote\svthefootnote%
}

\begin{document}
	
\title{Talk to Me: A Case Study on Coordinating Expertise in Large-Scale Scientific Software Projects}
\author{\IEEEauthorblockN{Reed Milewicz and Elaine M. Raybourn}
\IEEEauthorblockA{Sandia National Laboratories, 1611 Innovation Pkwy SE, Albuquerque, New Mexico 87123}}
\maketitle

\begin{abstract}

Large-scale collaborative scientific software projects require more knowledge than any one person typically possesses. This makes coordination and communication of knowledge and expertise a key factor in creating and safeguarding software quality, without which we cannot have sustainable software. However, as researchers attempt to scale up the production of software, they are confronted by problems of awareness and understanding. This presents an opportunity to develop better practices and tools that directly address these challenges. To that end, we conducted a case study of developers of the Trilinos project. We surveyed the software development challenges addressed and show how those problems are connected with what they know and how they communicate. Based on these data, we provide a series of practicable recommendations, and outline a path forward for future research.

\end{abstract}


\blankfootnote{
Sandia National Laboratories is a multimission laboratory managed and operated by National Technology \& Engineering Solutions of Sandia, LLC, a wholly owned subsidiary of Honeywell International Inc., for the U.S. Department of Energy's National Nuclear Security Administration under contract DE-NA0003525.}

\section{Introduction}

Large-scale scientific software projects are among the most knowledge-intensive undertakings, consisting of extremely diverse communities of practice and inquiry. For example, a climate modeling application can consist of numerous codes for modeling the atmosphere and the ocean, each of which is written by a distinct research team. The effective realization of such an application in an high-performance computing (HPC) environment relies heavily upon people with backgrounds in computational science and software engineering. The orchestration of that talent demands disciplined project management and communication with stakeholders. Thousands of person-years of labor are poured into the software development over the course of decades. 

Given the long lifespan and criticality of these projects, \textbf{sustainability} has been a focal point of research in recent years. By sustainability, we mean the ability of the software to continue to function as intended in the future, which is necessary for the reliability and reproducibility of research~\cite{hettrick2016research}. Sustainability is a multi-faceted challenge that encompasses both social and technical aspects of software development. In this work, we focus on the social aspect: the creation, communication, and use of knowledge integral to the scientific software development process. Large scientific software projects require diverse forms of expertise, bringing together people of different backgrounds and perspectives; to have success, there must be close, effective interaction among those parties~\cite{cohen2014simplifying}. Unfortunately, as we attempt to scale up these projects, we are confronted by barriers -- logistical, technical, and cultural -- that make it hard for people to share and apply what they know. These challenges increase both the cost and difficulty of software development and maintenance which ultimately threatens sustainability. 

From a software engineering perspective, more work is needed to create better tools and methodologies to manage and maintain that software development knowledge. However, as Dennehy and Conboy observe, the culture and context of a software project are ``critical determinants of software development success'' and that ``a method, practice, or tool cannot be studied in isolation''~\cite{dennehy2016going}. For these reasons, we offer a survey and study of knowledge management practices within the Trilinos project, a keystone scientific software library at Sandia National Laboratories~\cite{heroux2005overview}. In order to identify targets for intervention, we model how knowledge is created and, shared and its relationship to common software development challenges.

\subsection{Motivating Example}

Robust public investment into next-generation supercomputers is vital to the scientific enterprise. At the same time, the enormous sums of money that must be spent to construct and maintain these tools make it incumbent on their users to be accountable to the taxpayers. For this reason, government agencies stipulate rigorous requirements that must be met both by the machine and the software that it runs; a supercomputer must provide sufficient capabilities and the software must be able to fully utilize them. In the acceptance testing phase of supercomputer acquisition and software utility, participating research organizations put forward representative codes to be run on a novel architecture, and code performance is then compared against the capabilities advertised by the vendor. 

In the past year, the government requirements were tested when an well-respected application powered by Trilinos struggled to scale beyond $2^{17}$ Message Passing Interface (MPI) processes during an acceptance phase, resulting in a nearly 30\% drop in performance on the target architecture. Although all other applications passed the acceptance test and the contract was completed successfully, the issue flagged a potential ``time bomb'' for numerous applications and had to be corrected~\cite{timebombreport}. A team of researchers was given several months to locate the bug. The issue was finally resolved by a Trilinos scientist-developer who volunteered three weeks of his time to uncover it. The ultimate cause of the issue was a bug with two causes. First and foremost, a Trilinos meshing package upon which applications depended misused an MPI function, \textit{MPI\_Reduce\_scatter}, due to a misunderstanding of its semantics. In addition to this, it was found that the vendor-supplied implementation of the function was inefficient, which contributed to the overall slowdown. The issue was fixed by splitting the call into separate calls to \textit{MPI\_Reduce} and \textit{MPI\_Scatter}.

The lead author in this work is embedded with the Trilinos team, and both authors are members of the Interoperable Design of Extreme-scale Application Software (IDEAS) project\footnote{https://ideas-productivity.org}, which focuses on improving the productivity and sustainability of scientific software projects. As part of that mission, we carried out a subsequent investigation into the incident which revealed a deeper mystery: the \textit{exact same bug} had been introduced, found, and fixed in Trilinos multiple times over the years. The offending code was first introduced in three packages between 1998-2000 and fixed in 2005, copied line-for-line into a fourth package in 2004 and fixed again in 2015, and finally introduced into the meshing package in 2014 and fixed in 2017. In each case, the discovery and solutions were socialized, comments were made in the code, and notes were left in an issue tracker, but that information did not flow to the right parties in each subsequent incident. 

We stress that none of this reflects poorly upon the project; Trilinos is the work of a preeminent, world-class team of researchers and developers. Rather, this presents an opportunity for us to better understand the challenges that large scientific software teams face. As Moe et al. describe, the hallmark of large-scale software is that no one can know everything~\cite{moe2014networking}. Good, strategic communication and organization of expertise are necessary for a team to reach its full potential. For scientific software developers, that leads to several pertinent questions. How do researchers share (or fail to share) their knowledge? What practices or policies could be enacted to prevent or mitigate problems like the ones we have described in our motivating example? These we formulate as specific research questions:

\begin{itemize}
	\item \textbf{RQ1}: Do scientific software developers face problems in sharing their knowledge?
	\item \textbf{RQ2}: How does individual and organizational knowledge affect those problems?
	\item \textbf{RQ3}: How is that knowledge communicated?
\end{itemize}

\section{Background}

\subsection{Terminology}

There is no single, agreed-upon definition for what constitutes \textbf{knowledge} in a project. In this work, we use the definition provided by Davenport and Prusak, who state that knowledge is ``a fluid mix of framed experience, values, contextual information, and expert insights that provides a framework for evaluating and incorporating new experiences and information. It originates in and is applied in the minds of knowers. In organizations, it often becomes embedded not only in documents or repositories but also in organizational routines, processes, practices, and norms''~\cite{davenport1998working}. For example, Parise et al. show how a senior research scientist at a company is valuable not only for their own expertise but also for their ``critical relationships'' with other knowledgeable people (e.g., in academia)~\cite{parise2006strategies}. In other words, a successful research project must exercise both individual knowledge and those individuals' connections to other sources of knowledge.

We adopt the refined model introduced by Kelly, one built upon a decade of invaluable studies of scientific software development that better captures this dynamic and model the communication environment of our case study~\cite{kelly2015scientific}. It consists of five components or knowledge domains: \textbf{real world} (the phenomena being studied), \textbf{theory-based} (the models used to understand those phenomena), \textbf{software} (the development conventions and practices), \textbf{execution} (the tools and environment needed to create and run the software), and \textbf{operational} (the relationship between the use of the software solution and the real world problem). Each of these elements both inform the solution and many drive each other; for example, there is feedback between the theory and the real world, between writing the software and executing it on the hardware, and between the use of the software and its application to the real world problem. We used this model in developing our survey and in interpreting our findings.

\subsection{Knowledge in Large-scale Software Projects}

The phenomena we study in this work are based on those in LaToza et al., a study of the work habits and mental models of software developers~\cite{latoza2006maintaining}. In that work, the authors found that developers go to great lengths to maintain a tacit mental model of their software project, one that is reinforced through face-to-face communication with others and clear delineation and awareness of responsibilities. These needs become more difficult to satisfy as the size and scope of software projects grow. Scaling up software development methodologies that emphasize close communication and coordination (e.g., agile methodologies) is currently an open area of research~\cite{dingsoyr2014towards}~\cite{rolland2016problematizing}. This is especially pertinent for scientific software projects which often follow agile-like development methods~\cite{sletholt2012we}. The available evidence suggests that knowledge sharing at scale requires intentional practices that are tailored to organizational culture~\cite{santos2015fostering}, but it is not clear just what that means for scientific software. 

\subsection{Scientific Software Culture}

Studies of complex R\&D projects in industry have shown that a lack of a common language creates barriers to the communication and codification of knowledge generated by project activities~\cite{santos2012knowledge}. Because of the intimate relationship between software and science, the development process requires a diverse assortment of both domain experts and software engineers~\cite{mesh2013scientific}, and these teams are frequently distributed and multi-organizational~\cite{heaton2015claims}. As a multidisciplinary endeavor, each member of a project can have highly specialized knowledge that isn't easily transferred from one person to another. This is also a frequent source of conflict, such as between scientists and software engineers~\cite{segal2008scientists}~\cite{kelly2007software}, as well as between scientists from different disciplines~\cite{hara2003emerging}~\cite{jakobsen2004barriers}.

Scientists typically see software as a tool for creating and expanding scientific understanding, and less emphasis is placed on activities which concern knowledge of the software itself, such as planning or documentation~\cite{segal2008scientists}. There is often an implicit assumption that a scientific model and its implementation in code are connected such that knowledge of one translates to the other. This leads scientists to use software that they don't truly understand and write software without creating artifacts needed to understand it~\cite{hinsen2015technical}. Additionally, within the scientific community at large, there is a drive to produce publishable research, as publications are a pathway to funding, positions, and prestige~\cite{anderson2007perverse}. Studies have shown that scientists, when placed under pressure to publish, tend to focus on activities that lead to publications while neglecting those which do not~\cite{van2012intended}.

In many respects, such competing vocabularies, methods and agendas are not unique to the scientific software domain: the success of distributed and multidisciplinary teams almost never happens by accident. In the words of Ratcheva, ``simply putting people together in groups, representing many disciplines, does not necessarily guarantee the development of a shared understanding''~\cite{ratcheva2009integrating}.

\section{Case Study}

We surveyed developers of the Trilinos mathematical library project at Sandia National Laboratories to investigate the research questions presented in Section 1. Trilinos is a confederation of several dozen semi-independent packages. While packages may differ from one another in purpose, size, maturity, testedness, clients, and development teams, they are generally interoperable with one another, sharing datatypes, standardized interfaces, and a common vision for the project's ecosystem. It is rare for a client to use the entire codebase, rather they select a subset of the packages that pertain to their application. This leads to a combinatorial explosion in the number of ways in which Trilinos packages can be configured, built, and arranged. To give some perspective on the scale and complexity of the software, we characterize Trilinos in the context of other, similar DOE projects on Table~\ref{tbl:description}. The project is available as open-source software, hosted on GitHub\footnote{https://github.com/trilinos/Trilinos}, and follows a master-development branch model in which contributions are promoted to master if and only if all tests pass.

\begin{table}[]
\renewcommand{\arraystretch}{1.3}
\centering
\caption{Characterization of Trilinos compared with findings of Post and Cook 2000~\cite{post2000comparison}}
\label{tbl:description}
\begin{tabular}{ll}
Property         & Typical DOE CSE Project                                 \\ \hline
Code Complexity  & 20-50 independent packages                          \\
Code Size        & 500,000 LOC                                         \\
Project Age      & 10-35 years                                         \\
Release Schedule & 1-2 major releases, 20-100 minor releases per year. \\
Size of Teams    & 3-25 professionals                                  \\ \hline
                 & Trilinos                                              \\ \hline
Code Complexity  & 57 packages                                         \\
Code Size        & 2,247,210 SLOC									   \\
Project Age      & 17 years                                            \\
Release Schedule & 2 major releases per year                           \\
Size of Teams    & 3-6 professionals
\end{tabular}
\end{table}

We recruited participants for the survey through an internal developers' mailing list and print advertisements. The survey was distributed as a PDF file and as printed copies, and respondents provided code names to anonymize their identities. The mailing list consisted of 60 individuals and, among these, 29 developers were considered to be active, primary contributors. We received 36 responses, with 26 of which derived from the ``primary'' group, 5 from more peripheral developers on the mailing list, and an additional 6 responses from (mostly junior) members who were not subscribed to the list. This gives us a confidence level of 95\% with an interval of $\pm 6\%$ if we only consider the primary developers and $\pm 11\%$ for the entire population.

\begin{figure}
	\begin{center}
	\includegraphics[width=0.8\linewidth]{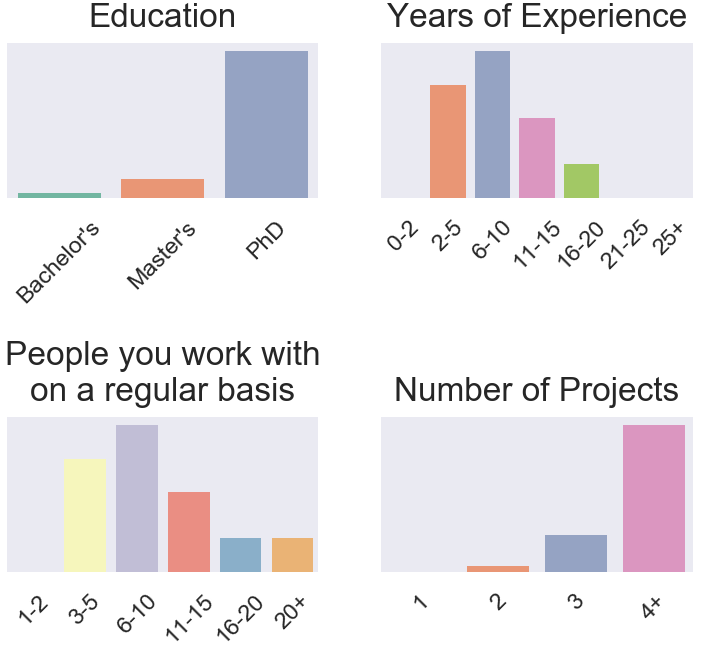}
	\end{center}
	\caption{A visual summary of the demographic information collected by our survey.}
	\label{fig:demographics}
\end{figure}

Our questionnaire consisted of multiple sections: demographic information, career priorities, methods of communication, areas of expertise, and problems encountered in software development. Additionally, we requested the GitHub handle of our respondents (removed from our published dataset) so that we could cross-reference the survey results with metrics on software contributions. We begin by presenting the demographic information, which is summarized in Figure \ref{fig:demographics}. The following are the highlights from this section:

\begin{itemize}
	\item 86\% of respondents had completed their PhD, and 83\% were members of staff (the remainder being interns, contractors, and postdocs, etc.). Collectively, respondents had between 233 to 345 years of combined experience, with the median respondent having from 11 to 15 years of experience.
	\item 77\% of respondents reported that they worked on 4 or more projects simultaneously, which typically means that they spend half their time working on a primary project, with the other half divided between three or more smaller, focused initatives. In total, 72\% of respondents reported working with 6 or more people on a regular basis, with the median respondent working with between 6 and 10 people.
	\item The research interests among Trilinos scientist-developers are very diverse, and, while there is some overlap, none of the respondents listed shared all the same interests. As a rule, projects like Trilinos do not hire for redundancy; each project member contributes unique skills to the project. 
\end{itemize}

In other words, our respondents tend to be highly educated, uniquely qualified, and in regular contact with a small fraction of the overall Trilinos team.


\section{Analysis}

In this section, we present the findings of our survey. Our presentation centers on commonly encountered software development issues and their relationship to individual and organizational knowledge. In Section \ref{sec:discussion}, we provide a discussion of the findings. Where correlations are provided, we use Pearson's $r$ and set $\alpha=0.05$ as the threshold for p-values to reject the null hypothesis. Survey materials and anonymized survey responses can be found online\footnote{https://github.com/rmmilewi/KnowledgeManagementSurvey}.

\begin{table}[]
\centering
\caption{A summary of the findings of our analysis.}
\label{tbl:findings}
\begin{tabular}{p{0.45\linewidth}p{0.45\linewidth}}
\rowcolor{gray!50}
Research Questions & Findings \\ \hline
\rowcolor{white}
\multicolumn{1}{p{0.45\linewidth}|}{\textbf{RQ1.} Do scientific software developers face problems in sharing their knowledge?} & The majority of respondents agreed with 13 out of 19 of our proposed problems (see Table~\ref{tbl:problems}). We found that position in the organizational network was moderately correlated with the total number of problems they reported, which is to say that well-connected people reported fewer problems. \\
\rowcolor{gray!25}
\multicolumn{1}{p{0.45\linewidth}|}{\textbf{RQ2.} How does individual and organizational knowledge affect those problems?} & We found weak correlations between operational and execution domain knowledge and four of the problems we studied. We also observed that ten problems are weakly correlated with having access to other people in specific areas of knowledge. \\
\rowcolor{white}
\multicolumn{1}{p{0.45\linewidth}|}{\textbf{RQ3.} How is that knowledge communicated?} & Communication strategies appear to affect eight problems, most notably in the case of face-to-face communication and expertise-finding problems. Additionally, the frequency and variety of communication correlates positively with self-perception of knowledge.
\end{tabular}
\end{table}

\subsection{Defining the Problems}

Recall the research questions presented in section 1 and that our RQ1 study question focused on identifying challenges, or problems, in sharing knowledge. For \textbf{RQ1}, respondents were presented with a list of commonly encountered issues in software development based on those in LaToza et al.~\cite{latoza2006maintaining}, a study of the work habits and mental models of software developers. For each, participants reported whether the issue was not a problem, a moderately difficult problem, or a difficult problem. The overall results can be seen in Table \ref{tbl:problems}. The median respondent reported having eleven of the nineteen problems, two of which were considered especially difficult. Our survey results suggest a strong consensus on the problems we listed, with majority support for 13 of the 19 problems.

\begin{table}[]
\centering
\caption{Respondent ratings of proposed problems. In the survey, problems were presented without headings and in a different order.}
\label{tbl:problems}
\begin{tabular}{p{5cm}p{3cm}}
\rowcolor{gray!50}
\multicolumn{1}{c}{Problem}                                                                 & \multicolumn{1}{p{3cm}}{\begin{tabular}[p{3cm}]{@{}p{3cm}@{}}This is a problem (\% agree)\\ / a difficult problem (\% agree)\end{tabular}} \\ \hline

\rowcolor{gray!50}
\textbf{Code Understanding}                                          & \\
\rowcolor{white}

Understanding the rationale behind a piece of code ($p_{coderationale}$)                                          & 63.9\% (13.9\%)                                                                                                             \\
\rowcolor{gray!25}
Understanding code that someone else wrote ($p_{otherscode}$)                                                  & 83.3\% (22.2\%)                                                                                                             \\
\rowcolor{white}
Understanding the history of a piece of code ($p_{codehistory}$)                                                & 58.3\% (8.3\%)                                                                                                              \\
\rowcolor{gray!25}
Understanding code that I wrote a while ago     ($p_{youroldcode}$)                                             & 22.2\% (0.0\%)                                                                                                              \\
\rowcolor{gray!50}
\textbf{Task Switching}                                          & \\
\rowcolor{white}

Having to switch tasks often because of requests from my teammates or manager ($p_{taskrequest}$)               & 75.0\% (38.9\%)                                                                                                              \\
\rowcolor{gray!25}
Having to switch tasks because my current task gets blocked  ($p_{taskblocked}$)                                & 55.6\% (11.1\%)                                                                                                             \\
\rowcolor{white}
Having to divide my attention between many different projects   ($p_{dividedattention}$)                             & 94.4\% (58.3\%)                                                                                                             \\
\rowcolor{gray!50}
\textbf{Modularity}                                          & \\
\rowcolor{white}
Being aware of changes to code elsewhere that impact my code  ($p_{changeothers}$)                               & 58.3\% (11.1\%)                                                                                                             \\
\rowcolor{gray!25}
Understanding the impact of changes I make on code elsewhere  ($p_{changeself}$)                               & 61.1\% (2.8\%)                                                                                                              \\

\rowcolor{gray!50}
\textbf{Links Between Artifacts}                                          & \\
\rowcolor{white}

Finding all the places code has been duplicated ($p_{duplication}$)                                             & 58.3\% (2.8\%)                                                                                                              \\
\rowcolor{gray!25}
Understanding who ``owns'' a piece of code    ($p_{ownership}$)                                                 & 38.9\% (0.0\%)                                                                                                              \\
\rowcolor{white}
Finding the bugs related to a piece of code   ($p_{bugsincode}$)                                               & 75.0\% (8.3\%)                                                                                                              \\
\rowcolor{gray!25}
Finding code related to a bug   ($p_{bugrelatedcode}$)                                                             & 83.3\% (11.1\%)                                                                                                             \\
\rowcolor{white}
Finding out who is currently modifying a piece of code  ($p_{modifiers}$)                                     & 33.3\% (0.0\%)                                                                                                              \\
\rowcolor{gray!50}
\textbf{Team}                                          & \\
\rowcolor{white}

Convincing managers that I should spend time rearchitecting, refactoring, or rewriting code ($p_{convincingmanagers}$)  & 41.7\% (25.0\%)                                                                                                             \\
\rowcolor{gray!25}
Convincing developers to make changes to code I depend on  ($p_{convincingdevelopers}$)                                  & 61.1\% (16.7\%)
                                                                                        \\
\rowcolor{gray!50}
\textbf{Expertise Finding}                                          & \\
\rowcolor{white}

Finding the right person to talk to about a piece of code     ($p_{rightpersoncode}$)                               & 50.0\% (8.3\%)                                                                                                              \\
\rowcolor{gray!25}
Finding the right person to talk to about a bug      ($p_{rightpersonbug}$)                                        & 38.8\% (5.6\%)                                                                                                              \\
\rowcolor{white}
Finding the right person to review a change before a check-in    ($p_{rightpersonreview}$)                            & 25.0\% (5.6\%)                                                                                                              \\
\end{tabular}
\end{table}

The three most commonly reported problems were dividing attention between projects ($p_{dividedattention}$), understanding other's code ($p_{otherscode}$), and finding bugs related to code ($p_{bugrelatedcode}$). Given the widespread agreement on so many problems, we first want to test whether problems in each category are strongly correlated with one another, which would suggest that they measure the same construct; in other words, might our problems reflect a small number of common causes? We can infer this using Cronbach's alpha (which we will refer to as $C_{\alpha}$) as a measure of interrelatedness or reliability~\cite{cronbach1951coefficient}. The score ranges from 0 to 1, and the rule of thumb is that $C_{\alpha} >= 0.70$ suggests a set of items has good internal consistency; strong internal consistency in survey measures suggests that they all measure some common, latent variable. In Table \ref{tbl:cronbach}, we see that this holds true for only two of the five categories. From this, we can conclude that while many problems are common, they are often independent of each other and will need to be addressed separately (e.g., finding a reviewer for your code and finding someone to talk about a bug require different information).

\begin{figure}
	\includegraphics[width=0.45\textwidth]{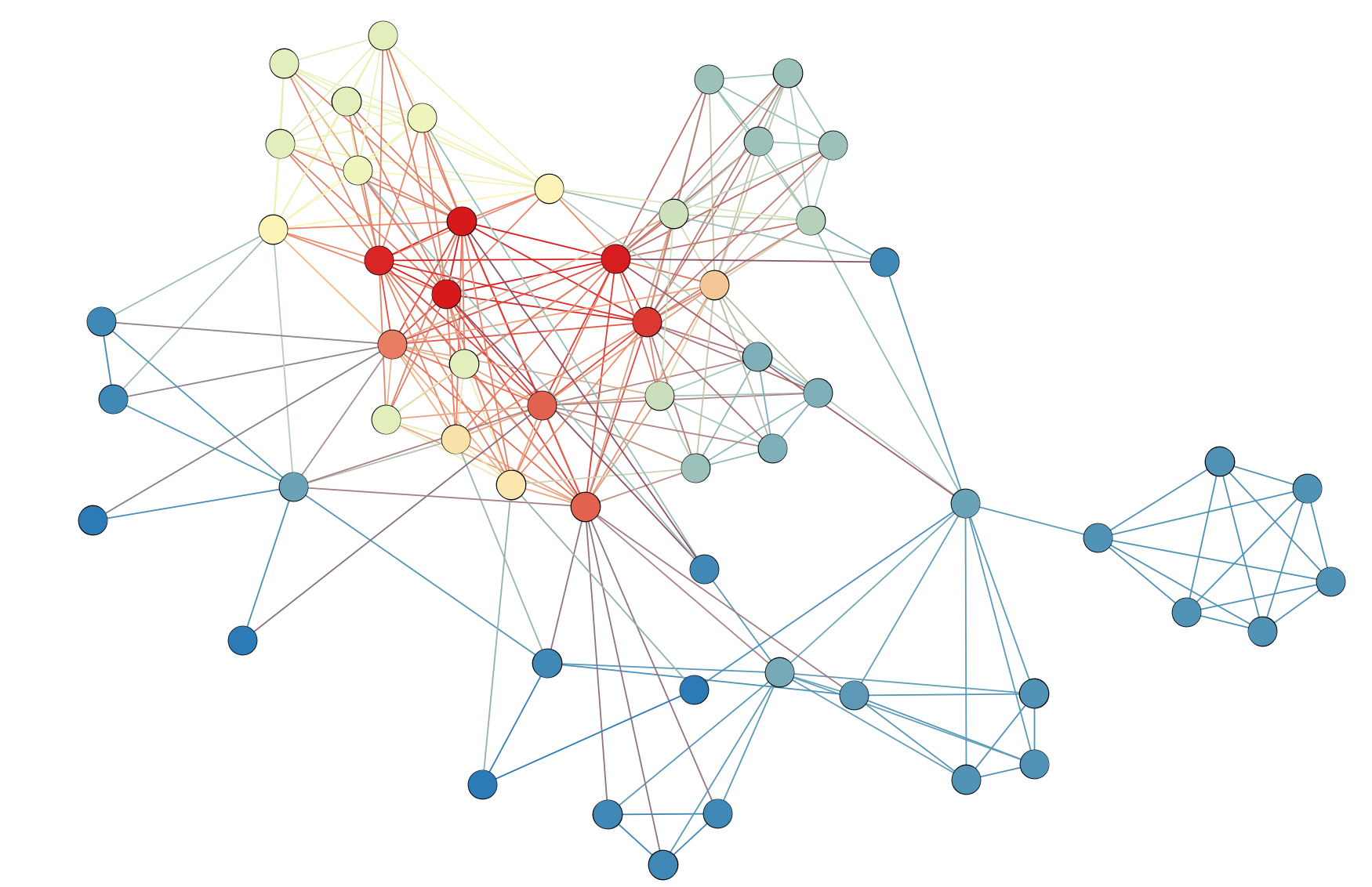}
	\caption{A graph of Trilinos developers on GitHub assigned to teams, where each edge indicates that two developers belong to the same development sub-team. The graph is color-coded to reflect the number of triangles formed between a node and its neighbors, red being more interconnected and blue being less interconnected.}
	\label{fig:teamgraph}
\end{figure}

\begin{figure}
	\includegraphics[width=1\linewidth]{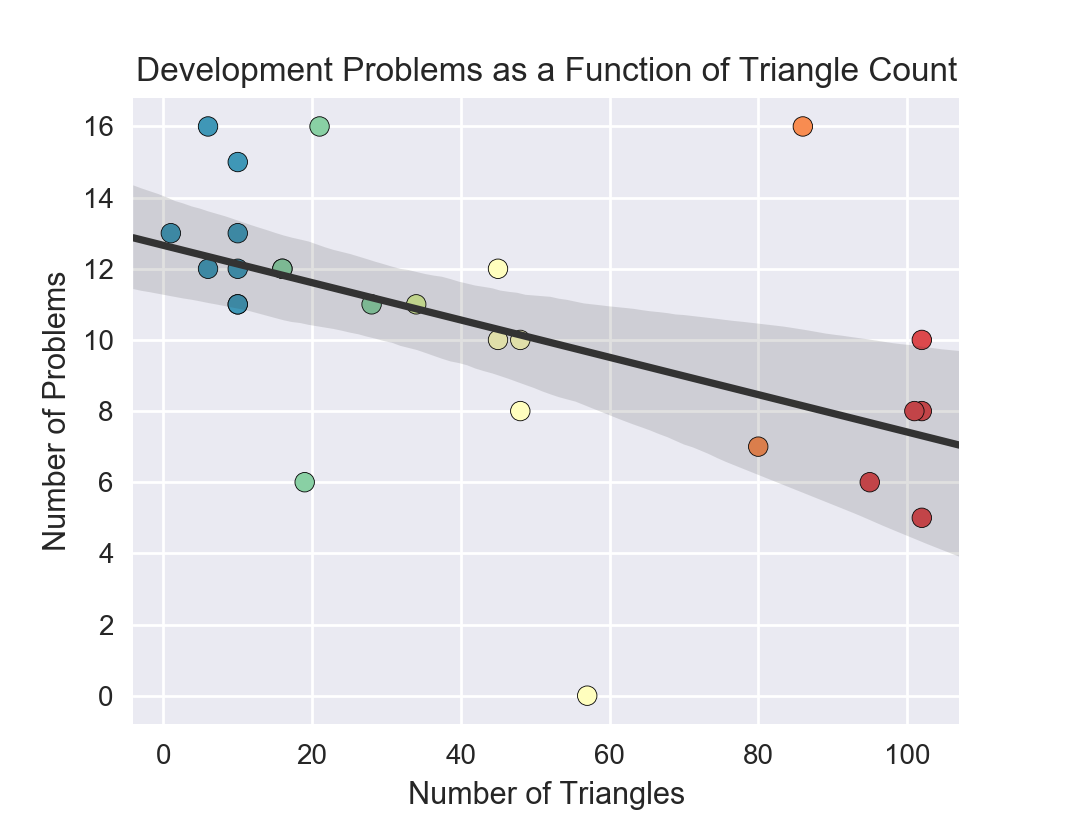}
	\caption{A linear regression on the number of problems reported by developers as a function of their triangle count  ($r=-0.434$,$p=0.008$). Even without considering our survey data, the plot demonstrates how problems of understanding and awareness are correlated with the way in which developers are situated in the organization.}
	\label{fig:triangleregplot}
\end{figure}

Likewise, we also asked whether these problems could be explained by simple demographic or organizational measures (e.g., do more experienced people have fewer problems), without our nuanced survey data. For this, we examined the relationship between problems and organizational network structure by looking at team composition. The Team API of GitHub allows projects to group developers into teams, and the Trilinos project uses this feature to match developers with particular packages or cross-cutting concerns (e.g., the framework team); we found 57 teams, one of which was a global team of all developers that we excluded from our analysis. From this data we produced the team member graph seen in Figure \ref{fig:teamgraph}, which provides a rough estimation of the lines of communication between developers. For each node in the graph, we computed its triangle count, which is the number of triangles (cyclic paths of length 3) formed between it and its neighbors; triangle counting is a common measure in social network analysis and it is the underpinning for measures such as the clustering coefficient (see ~\cite{suri2011counting}). As we illustrate in Figure \ref{fig:triangleregplot}, there is a moderate inverse correlation between triangles and problems: the more embedded a person is in the team network, the less likely they are to have problems. Additionally, the triangle count correlates with three specific problems at the $\alpha=0.05$ level: $p_{youroldcode}$ ($r=-0.335,p=0.045$), $p_{bugsincode}$ ($r=-0.465,p=0.004$), and $p_{rightpersonbug}$ ($r=-0.353,p=0.004$).

\begin{table}[]
\centering
\caption{Cronbach's alpha scores for problem categories.}
\label{tbl:cronbach}
\begin{tabular}{ll}
\rowcolor{gray!50}
Category           & $C_{\alpha}$ \\  \hline
\rowcolor{gray!25}
Code Understanding & 0.770        \\
\rowcolor{white}
Task Switching     & 0.715        \\
\rowcolor{gray!25}
Modularity         & 0.474        \\
\rowcolor{white}
Artifacts          & 0.595        \\
\rowcolor{gray!25}
Team               & 0.594        \\
\rowcolor{white}
Expertise Finding  & 0.579
\end{tabular}
\end{table}

Our next question is whether this embeddedness is merely a proxy of any of our other demographic variables. We see that embeddedness is neither a function of experience or seniority ($r=-0.015$,$p=0.932$) nor the number of projects ($r=0.107$,$p=0.533$) nor the total number of people that people regularly work with ($r=-0.242$,$p=0.155$). Moreover, none of these demographic measures can directly account for the number of problems that people face. We do find three problems that have weak correlations with these factors: $p_{bugrelatedcode}$ ($r=-0.350$,$p=0.036$), and the number of projects and both $p_{codehistory}$ ($r=0.361$,$p=0.030$) and $p_{changeothers}$ ($r=-0.342$,$p=0.041$). However, even though many of the problems we listed relate to understanding and awareness, raw quantities of experience and contact with others do not correlate with the overall number of problems people face. From this, we can conclude that we need to dig deeper into what and who people know, how they know them, and why.

\textbf{Summary}: All of the problems we investigated are, in some sense, problems of coordination, understanding, and awareness; they concern what people know, who they know, and how they use that information. These challenges are certainly not unique to scientific software, but they take on added weight and meaning given the demanding and knowledge-intensive nature of the work. Our analysis of the survey data suggests that these problems have multiple underlying drivers that do not neatly align with our categories. The closest we can come to a unifying explanation is that the number of problems respondents have is moderately correlated with embeddedness in the organizational network. This is not altogether surprising: studies of R\&D organizations have often drawn attention to the value of network centrality in amplifying an individual's impact and increasing their access to knowledge (see ~\cite{nerkar2005evolution}). 

\subsection{What and Who Do They Know?}

For \textbf{RQ2}, we want to characterize the range of expertise of each participant and their access to others with expertise. We selected eight topics corresponding to the five knowledge areas described in Kelly 2015~\cite{kelly2015scientific}, and these are described in Table \ref{tbl:knowledge}. For each topic, we asked participants to provide a self-assessment of their own familiarity with the topic on a 5-point Likert scale ranging from ``not very knowledgeable'' to ``very knowledgeable''. Additionally, we asked participants whether they worked with someone else that they ``could turn to for help on that topic''. Our survey results can be seen in Table \ref{tbl:knowledge}.

\begin{table*}[]
\centering
\caption{Results for knowledge self-assessment questions}
\label{tbl:knowledge}
\resizebox{\textwidth}{!}{%
\begin{tabular}{llllll}
\multicolumn{1}{c}{Topic} & \multicolumn{1}{c}{Knowledge Area} & \multicolumn{1}{c}{Histogram} & \multicolumn{1}{c}{Median (out of 5)} & \multicolumn{1}{c}{\% know someone else} &  \\ \cline{1-5}
Knowledge of the real-world phenomena that the software is used to study. & Real-World & \knowledgespark{0.02}{0.2}{0.34}{0.23}{0.2} & 3 & 63.8\% &  \\
The selection of mathematical techniques to attack a problem. & Theory & \knowledgespark{0.03}{0.118}{0.118}{0.471}{0.265} & 4 & 63.8\% &  \\
Software design & Software & \knowledgespark{0.03}{0.03}{0.33}{0.38}{0.22} & 4 & 50.0\% &  \\
Software construction & Software & \knowledgespark{0}{0.28}{0.11}{0.42}{0.44} & 5 & 50.0\% &  \\
Compilers and compiler optimizations & Execution & \knowledgespark{0.14}{0.17}{0.22}{0.39}{0.8} & 4 & 55.0\% &  \\
The effects of hardware architecture on algorithm performance & Execution & \knowledgespark{0.11}{0.14}{0.33}{0.25}{0.17} & 3 & 58.3\% &  \\
Using a version control system & Execution & \knowledgespark{0}{0.05}{0.11}{0.36}{0.47} & 5 & 55.0\% &  \\
How the software is integrated with client codes & Operational & \knowledgespark{0.05}{0.16}{0.14}{0.36}{0.27} & 4 & 47.2\% &
\end{tabular}
}
\end{table*}

\begin{figure}
\includegraphics[width=1\linewidth]{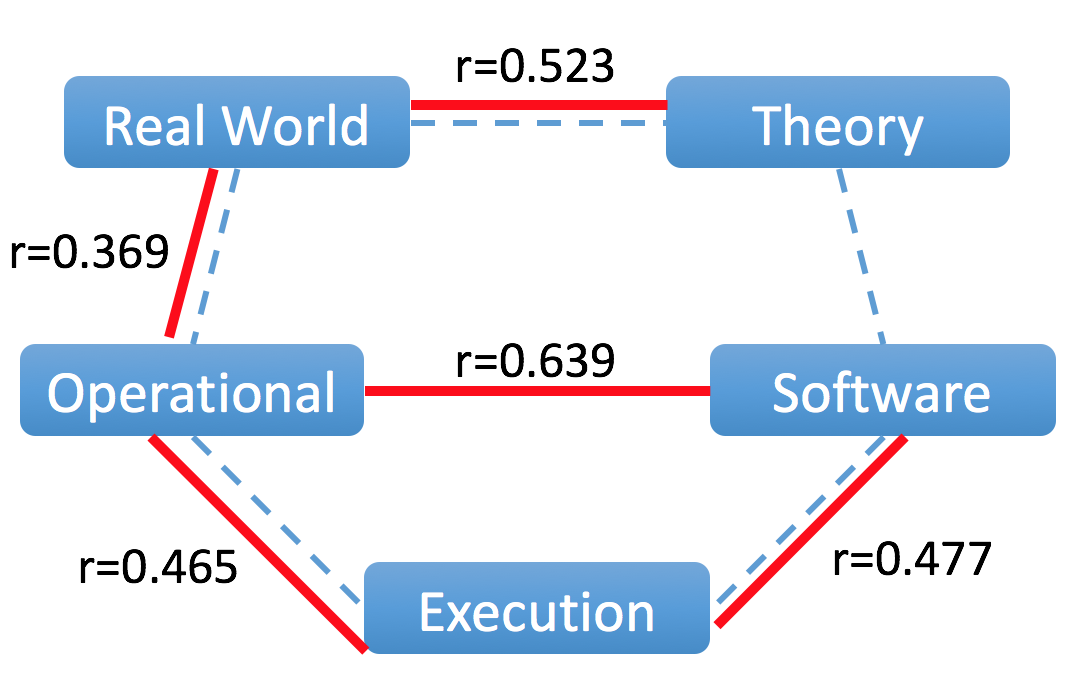}
\caption{A map of correlations between knowledge categories significant at the $\alpha=0.05$ level. Dashed blue lines indicate relationships predicted by the model, and solid red lines indicate relationships suggested by our data.}
\label{fig:kelly}
\end{figure}

As a litmus test for our topic choices, we compare our survey findings against the Kelly five factor model introduced in section II by aggregating measures according to category, as can be seen in Figure \ref{fig:kelly}. Our findings lend strong support to the model with agreement on eight out of ten possible edges. We found a moderate relationship between operational and software knowledge ($r=0.639$,$p=0.00004$) which is not predicted by the five factor model; this is likely an artifact of the Trilinos team being library developers (i.e. writing code for other people's code). We also note a missing edge between theory and software knowledge: many Trilinos developers translate theory into code, but there's no support for the notion that one domain is used to increase knowledge in the other. If we dig into the data, we find that scores increase with years of experience for every topic \textit{except} for mathematics ($r=0.001$,$p=0.994$). This suggests that self-perception of knowledgeability in this area is relatively fixed, so our survey instrument may not be picking up on any cross-pollination that may happen between theory domain and software domain topics.

We found knowledge had a weak inverse correlation with three of the of the nineteen problems and a moderate inverse correlation with one. First, finding the right person to talk about a piece of code ($p_{rightpersoncode}$) was less likely a problem among respondents with knowledge of version control ($r=-0.337$,$p=0.044$) and hardware ($r=-0.348$,$p=0.037$). Second, finding duplicated code ($p_{duplication}$) was seen as easier by respondents with high knowledge of compilers ($r=-0.362$,$p=0.030$) and hardware ($r=-0.336$,$p=0.045$). Third, knowledge of client codes and of version control were related to the problem of understanding code written by others ($p_{otherscode}$, $r=-0.344$,$p=0.040$ and $r=-0.365$,$p=0.040$). Finally, compiler expertise has a moderate correlation with the problem of determining code ownership ($p_{ownership}$, $r=-0.503$,$p=0.002$). Taken all together, this suggests that the common driver in these findings is a deep awareness of the work context, such as the needs of clients or the execution of the code on target architectures. Given our findings and sample size, more research is warranted.

We also found that people with a high knowledge of design were more likely to report problems with receiving requests to switch tasks ($r=0.353$,$p=0.035$) and having to divide their attention between projects ($r=0.337$,$p=0.043$). Additionally, people with a high knowledge of math were more likely to have problems with tracking which people are modifying different pieces of code ($r=0.358$,$p=0.037$).


Next, we wanted to know what benefits there are to being connected with other people who have knowledge. We carefully worded the prompt to check for contacts that a respondent ``could turn to for help'', with the expectation that people who have problems are more likely to seek out people who can help them. The average respondent reported having contacts that covered 4.4 of the 8 topics, and the amount of coverage was found to be moderately correlated with the number of problems ($r=0.455$,$p=0.005$). Additionally, the only indicator that someone had a contact for one topic was that they had a contact for another topic (average $r=0.808$, average $p=0.0000002$). This is to say that the ability or tendency to engage in this kind of networking was independent of an individual's background or position in the team.

Ten of the nineteen problems are weakly correlated with having people to reach out to (average $r=0.389$, average $p=0.0245$) spread across the following categories: code understanding ($p_{otherscode}$,$p_{codehistory}$, and $p_{youroldcode}$), task switching ($p_{taskrequest}$,$p_{taskblocked}$, and $p_{dividedattention}$), modularity ($p_{changeothers}$), links between artifacts ($p_{ownership}$ and $p_{bugsincode}$), and team  ($p_{convincingdevelopers}$). This may suggest that people on the Trilinos team are motivated to seek out knowledge-related contacts in order to maintain awareness of code and to negotiate and coordinate with others. 


\textbf{Summary}: Knowledge in multiple domains is critical at every stage of the scientific software lifecycle. However, while the problems we researched in this paper pertain to scientific software development, they are not problems solved by writing better software or doing better research. Recall the model proposed by Kelly in section II: real-world, theory, and software domain knowledge provided no measurable benefit. On the other hand, operational and execution domain knowledge were correlated with four of the nineteen problems; the common denominator is a deep awareness of how the software is assembled and used --- knowledge which helps bridge the gap between different domains of activity. Finally, we found that the majority of problems were correlated with seeking out help from others. For example, people who had problems understanding other's code were more likely to have a contact knowledgeable in software construction ($r=0.447$,$p=0.006$).


\subsection{How do they Communicate?}

\begin{figure}
\includegraphics[width=1\linewidth]{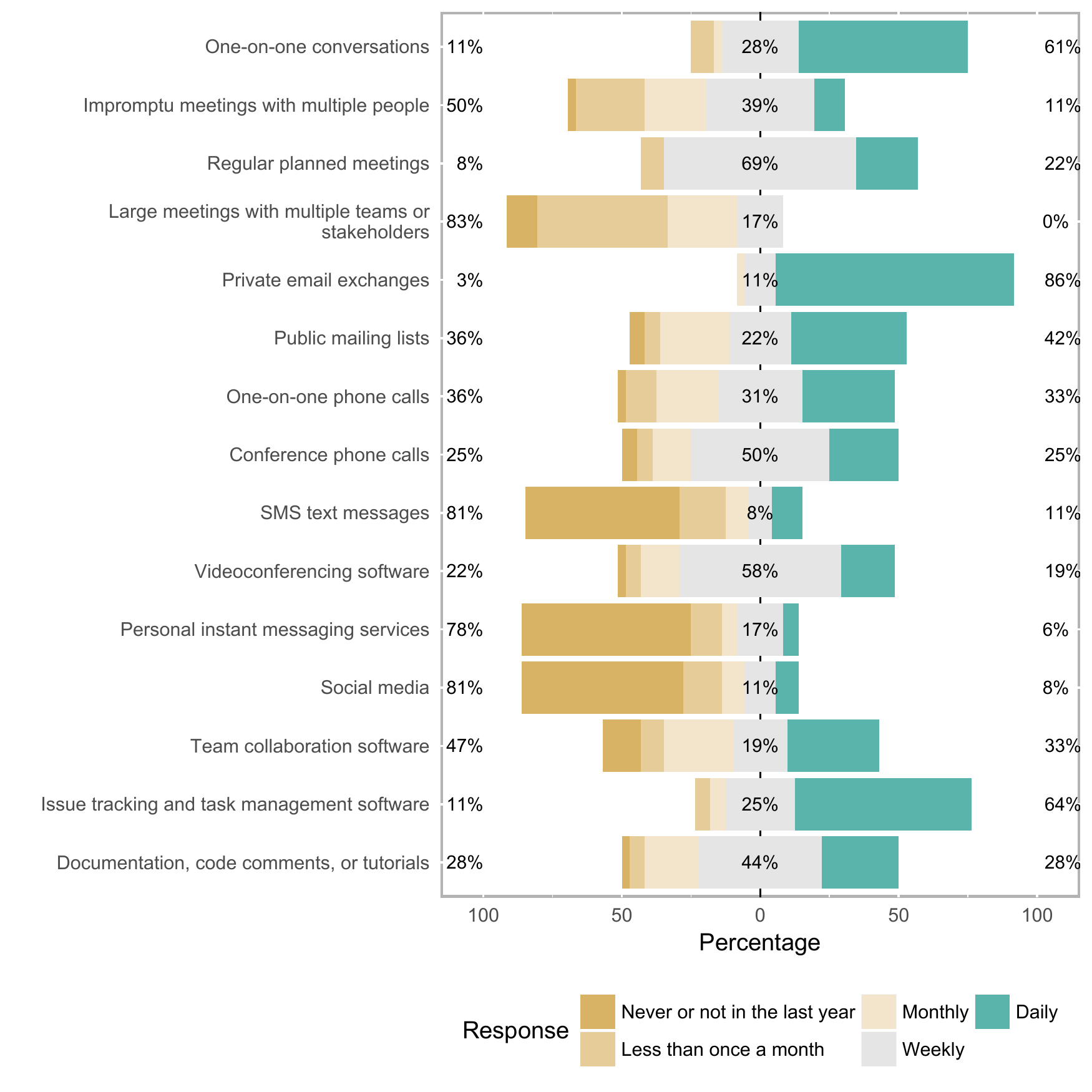}
\caption{The results of the Communications portion of the survey, which examined how and how frequently developers communicated with each other.}
\label{fig:commfreq}
\end{figure}

Finally, we want to know how expertise is communicated among Trilinos developers, and how it is correlated with other survey questions. We provided respondents with a list of different communication media, and for each we asked them to describe how frequently they used them on a 5-point scale from ``never or not in the last year'' to ``daily''. The results can be seen in Figure \ref{fig:commfreq}. We found that knowledge scores were moderately correlated to communication scores in that those who communicated more frequently considered themselves more knowledgeable and vice versa (for average scores, $r=0.491,p=0.002$). Meanwhile, eight of the nineteen problems were correlated with differences in communication strategies.

Face-to-face communication was a contributor to all problems in the expertise finding category. Finding a reviewer for code ($p_{rightpersonreview}$) was inversely correlated with frequent one-on-one meetings ($r=-0.360$,$p=0.031$); a similar relationship was found for one-on-one phone calls and finding someone to talk about a bug ($p_{rightpersonbug}$,$r=-0.403$,$p=0.016$). Unstructured meetings with multiple people, meanwhile, was implicated in finding the right person to talk about a bug ($r=-0.360$,$p=0.043$). Finally, large meetings with multiple stakeholders were correlated with both finding people to talk about code ($p_{rightpersoncode}$,$r=-0.383$,$p=0.021$) as well as bugs ($r=-0.339$,$p=0.031$). The takeaway is that these problems are a function of close and sustained communication, moreso than mere awareness of others.

Meanwhile, we found that digital communications were effective at solving some challenges while fueling others. Private email was correlated with understanding of how other people's distant changes may affect your code ($p_{changeothers}$,$r=-0.372$,$p=0.430$), but it is also a high-bandwidth channel of communication correlated with divided attention problems ($p_{dividedattention}$,$r=0.437$,$p=0.008$). Likewise, video conferencing has made it possible to keep people involved in many different projects, even when separated over great distances, but it also is correlated with divided attention ($r=0.451$,$p=0.006$). Meanwhile, team collaboration software (e.g., Jira, Confluence, etc.) are inversely correlated with the problem of knowing who responsible for work items ($p_{ownership}$,$r=-0.329$,$p=0.50$) as well as piecing together the origins of bugs ($p_{bugrelatedcode}$,$r=-0.383$,$p=0.021$), but unaccounted slack time which could be spent on rearchitecting and refactoring activities may be harder to come by ($p_{convincingmanagers}$,$r=0.401$,$p=0.015$). Lastly, we found that use of documentation was inversely correlated with the problem of bug finding ($p_{bugrelatedcode}$,$r=-0.396$,$p=0.016$).

\textbf{Summary}: Our analysis suggests that quantity and quality of communication are correlated with eight of the nineteen problems. Frequent face-to-face communications may enable expertise-finding activities, implying a need for ongoing, sustained contact. Next, digital communication strategies are useful for protecting modularity and understanding the links between artifacts, but the communication overhead also introduces new challenges (e.g., dealing with divided attention). Finally, knowledge assessments seem to mirror communication assessments; those who communicate more know more, or at the very least are more confident in their knowledge.

\section{Discussion}
\label{sec:discussion}

As we attempt to scale up the production of scientific software to meet the demands for innovation, we are beset by problems of understanding, coordination, and awareness. These are complex and multi-faceted issues for which there can be no single solution. In our case study, we found that the number of problems that respondents reported was, in some sense, a reflection of their ``embeddedness'' in the team, which motivated further investigation into how expertise is situated and accessed.


The most useful forms of expertise were those that allowed respondents to position themselves between domains of activity, namely between the code and the machine (execution knowledge) and between the developers and the clients (operational knowledge). The benefit of that knowledge is indirect. For example, respondents who know more about compilers had fewer problems identifying code ownership, but this is likely because compiler experts place more importance on tracking the sources of different codes. Likewise, people with a high knowledge of design had more attention problems, and this is probably because design work supports and coordinates the activities of many different people. Other relationships that fall out of the data, such as between knowledge of math and tracking code modifications may have more complex etiologies; participants who place a stronger focus on research and skills that support research may be devoting less time and energy towards maintaining awareness of the software project.

Most of the problems addressed in this study are not able to be solved by individuals in isolation, a fact that motivates respondents to seek out contacts across different areas of expertise. This was especially important for people having to negotiate and coordinate with others to carry out work and for those trying to maintain awareness of code written by others. Seeking out help did not reduce the frequency with which respondents reported problems, which implies that this is a risk mitigation rather than a risk reduction strategy. We infer that mere awareness and/or periodic contact is not enough to reduce the occurrence of issues: resolution of several of the problems we studied depended upon the quantity and quality of contact with others. Frequent communication had a positive correlation with knowledge across the board. With respect to particular problems, frequent face-to-face communications were important for locating and using other people's expertise. Meanwhile, digital communications were found to help with change awareness and bug-finding, but also had the potential to exacerbate attention problems and create new bureaucratic barriers.

\begin{table}[]
\centering
\caption{The number of problems in each category that have a statistically significant relationship with the items studied in our survey.}
\label{my-label}
\begin{tabular}{p{2cm}p{1.1cm}p{1cm}p{1cm}p{1cm}}
\rowcolor{gray!50}
Problem Area & Back-ground & What They Know & Knowing Who Knows & How They Communicate \\ \hline
\rowcolor{white}
\multicolumn{1}{p{2cm}|}{Code Understanding} &  (2/4) &  (1/4) &  (3/4) &  \\
\rowcolor{gray!25}
\multicolumn{1}{p{2cm}|}{Switching Tasks} &  &  (2/3) &  (3/3) &  (1/3) \\
\rowcolor{white}
\multicolumn{1}{p{2cm}|}{Modularity} &  (1/2) &  &  (1/2) &  (1/2) \\
\rowcolor{gray!25}
\multicolumn{1}{p{2cm}|}{Links Between Artifacts} &  (2/5) &  (3/5) &  (2/5) &  (2/5) \\
\rowcolor{white}
\multicolumn{1}{p{2cm}|}{Team} &  &  &  (1/2) &  (1/2) \\
\rowcolor{gray!25}
\multicolumn{1}{p{2cm}|}{Expertise Finding} &  (1/3) &  (1/3) &  &  (3/3)
\end{tabular}
\end{table}


\subsection{Recommendations}


The phenomena described in this paper are very common among large software development projects. However, not all solutions readily translate to the scientific software domain. Sletholt et al., a literature review on the use agile practices in scientific software development, found support for some agile methods but not others ~\cite{sletholt2012we}. The authors also caution that their evidence is strongest when considering ``small projects with relatively few team members.'' For example, in a project like Trilinos, where the number one complaint among developers is having too many projects and not enough time, daily stand-up meetings may not be a realistic solution for everyone. This being said, we have identified several well-supported solutions that we believe may be a good fit for large scientific software teams like Trilinos.

\textbf{Empowering knowledge brokers}: Boden et al. call attention to the role of knowledge brokers in distributed software development, that is, people capable of acting as bridges between different teams and domains of expertise; knowledge brokers are considered critical to enabling the flow of information between different sites~\cite{boden2009bridging}. Brokers tend to play an informal role in filling in structural holes in social networks, but works like Parise et al. 2006 argue that organizations should give formal recognition and power to these people~\cite{parise2006strategies}. In our case study, almost all of the Trilinos team is located within the same research building, but this is not a good guarantee of team cohesion: a recent study of R\&D organizations found that the frequency of scientific collaboration drops off given 100 feet of distance between offices~\cite{kabo2014proximity}. Along these lines, we note that 33\% of our respondents indicated that they knew no one they could turn to for help in any of the knowledge areas while having an average of 5 different problems that could potentially be mitigated by having useful contacts; this is a situation where brokers could be helpful. 

\textbf{Cultivating organizational awareness}: A benefit of having strong networks is the potential for serendipitous encounters. Santos et al. point out that the most effective knowledge sharing in complex R\&D projects takes place in bars after work~\cite{santos2012knowledge}. While strategies such as having knowledge brokers can help people locate specific expertise on demand, passive and casual exchanges of knowledge can clue people in to opportunities they might not have known about otherwise. This is echoed by Schossau and Wilson 2014, who found that one of the ``completely unanticipated'' benefits of Software Carpentry workshops was that they promoted awareness of technologies and methods, even if that information was not immediately useful~\cite{schossau2014sustainable}. It is possible to create these conditions through events such as interdepartmental lunches and seminars.

\textbf{Encouraging integrative work}: As our survey data show, quality (not just quantity) of interactions matters for problems such as expertise finding. This echoes the findings of Hara et al., who distinguished between complementary and integrative collaborations in research groups, the former requiring awareness and the latter requiring frequent, close communication~\cite{hara2003emerging}. In our case, we found that 36\% of respondents reported having no daily face-to-face interactions with other coworkers; the value of quiet isolation notwithstanding, there is also much to be said for close collaboration. However, the solutions in this category make greater demands on individuals. Pair programming, for instance, has been shown to have great potential in conventional software development, but it has seen only limited adoption among scientific software teams~\cite{sletholt2012we}. Another strategy commonly employed in industry is to occasionally rotate members between different teams in order to disseminate best practices~\cite{santos2015fostering}.

\section{Threats to Validity}

There are several potential threats to internal validity in this study. As with all surveys, our work is vulnerable to response biases. One concern in crafting this survey was social desirability bias, as our survey asks participants about their strengths and their weaknesses. This has come up in other surveys of scientist-developers such as Carver et al., which found that scientists tend to overestimate their own software development abilities~\cite{carver2013self}. We attempted to control for this by having a vetted protocol for collecting and storing survey data to protect the anonymity and confidentiality of responses; in general, we found that respondents were eager to volunteer information. Another concern was non-response bias because scientist-developers are notoriously preoccupied, but nevertheless we were able to get a sufficient number of responses. Moreover, as this was an organizational survey, we were able to precisely quantify the number of non-responses.

While our sample size is representative of the population, the population itself is a single team, and this raises questions about external validity. Most scientific software teams do not operate at the size and scale of Trilinos; a 2018 study by Pinto et al. found that 95\% of scientific software teams they surveyed had five members or fewer~\cite{pinto2018scientists}. However, large-scale projects (e.g., libraries) are foundational for the scientific software ecosystem, and the problems we studied are universal to large software projects regardless of domain. 

Lastly, we note that almost all of the correlations we report are weak to moderate in strength. This was not unexpected, as we were concerned with uncovering the relationships between software development problems and incidental, perhaps unintentional practices. Our findings are meant to spur further investigation into potential solutions that may tap into the dynamics of knowledge and communication we have described. 

\section{Related Work}

Szymczak et al. argue for a rational, document-driven approach to codifying the knowledge surrounding scientific software development, and introduces Drasil, a platform for accomplishing this~\cite{szymczak2016position}. Along these lines, Smith et al. provide a series of case studies on the application of document-driven design to scientific software~\cite{smith2016advantages}. We recognize the value of this approach, especially when it comes to improving usability and verifiability, but we caution that most software development knowledge is tacit and unable to be codified; many of the specific problems we have described in our work are not easily addressed by knowledge capture strategies.

On the subject of training and education, Gil et al. note that there is ``a very limited focus on issues of collaborative software development'' in the education of early-career scientists~\cite{gil2014collaborative}. Our work suggests that such training is of special importance to large-scale scientific software development.



\section{Conclusion}


In this work, we studied problems of communication and awareness in realizing large-scale scientific software by conducting a survey of developers of the Trilinos project, a key software library at Sandia National Laboratories. Scientific software development is vitally necessary to the modern scientific enterprise. However, achieving development at scale means confronting recurrent problems of awareness and understanding which pose risks to sustainability. Our takeaway is that several widespread development problems may be related to the way in which expertise is situated and communicated within the team, and presented several preliminary recommendations. We hope to test the validity of those recommendations by putting them into practice. Our findings underscore the need for more investigation into development methodologies suitable for large-scale scientific software development. To that end, our future work will include ethnographic research into the work practices of scientist-developers to further study the role of coordinating expertise and communication in large-scale scientific software development. 


\bibliographystyle{IEEEtran}
\bibliography{knowledge}

\begin{thebibliography}{10}
\providecommand{\url}[1]{#1}
\csname url@samestyle\endcsname
\providecommand{\newblock}{\relax}
\providecommand{\bibinfo}[2]{#2}
\providecommand{\BIBentrySTDinterwordspacing}{\spaceskip=0pt\relax}
\providecommand{\BIBentryALTinterwordstretchfactor}{4}
\providecommand{\BIBentryALTinterwordspacing}{\spaceskip=\fontdimen2\font plus
\BIBentryALTinterwordstretchfactor\fontdimen3\font minus
  \fontdimen4\font\relax}
\providecommand{\BIBforeignlanguage}[2]{{%
\expandafter\ifx\csname l@#1\endcsname\relax
\typeout{** WARNING: IEEEtran.bst: No hyphenation pattern has been}%
\typeout{** loaded for the language `#1'. Using the pattern for}%
\typeout{** the default language instead.}%
\else
\language=\csname l@#1\endcsname
\fi
#2}}
\providecommand{\BIBdecl}{\relax}
\BIBdecl

\bibitem{hettrick2016research}
S.~Hettrick, ``Research software sustainability: Report on a knowledge exchange
  workshop,'' 2016.

\bibitem{cohen2014simplifying}
J.~Cohen, C.~Cantwell, N.~C. Hong, D.~Moxey, M.~Illingworth, A.~Turner,
  J.~Darlington, and S.~Sherwin, ``Simplifying the development, use and
  sustainability of hpc software,'' \emph{Journal of Open Research Software},
  vol.~2, no.~1, 2014.

\bibitem{dennehy2016going}
D.~Dennehy and K.~Conboy, ``Going with the flow: An activity theory analysis of
  flow techniques in software development,'' \emph{Journal of Systems and
  Software}, 2016.

\bibitem{heroux2005overview}
M.~A. Heroux, R.~A. Bartlett, V.~E. Howle, R.~J. Hoekstra, J.~J. Hu, T.~G.
  Kolda, R.~B. Lehoucq, K.~R. Long, R.~P. Pawlowski, E.~T. Phipps
  \emph{et~al.}, ``An overview of the trilinos project,'' \emph{ACM
  Transactions on Mathematical Software (TOMS)}, vol.~31, no.~3, pp. 397--423,
  2005.

\bibitem{timebombreport}
K.~Agelastos, M.~Rajan, N.~Wichmann, P.~Lin, R.~Baker, S.~Domino, E.~Draeger,
  S.~Anderson, J.~Balma, S.~Behling, M.~Berry, P.~Carrier, M.~Davis,
  K.~McMahon, D.~Sandness, K.~Thomas, S.~Warren, and T.~Zhu, ``Performance on
  trinity phase 2 (a cray xc40 utilizing intel xeon phi processors) with
  acceptance-applications and benchmarks.''\hskip 1em plus 0.5em minus
  0.4em\relax Sandia National Laboratories, 2017.

\bibitem{moe2014networking}
N.~B. Moe, D.~{\v{S}}mite, A.~{\v{S}}{\=a}blis, A.-L. B{\"o}rjesson, and
  P.~Andr{\'e}asson, ``Networking in a large-scale distributed agile project,''
  in \emph{Proceedings of the 8th ACM/IEEE International Symposium on Empirical
  Software Engineering and Measurement}.\hskip 1em plus 0.5em minus 0.4em\relax
  ACM, 2014, p.~12.

\bibitem{davenport1998working}
T.~H. Davenport and L.~Prusak, \emph{Working knowledge: How organizations
  manage what they know}.\hskip 1em plus 0.5em minus 0.4em\relax Harvard
  Business Press, 1998.

\bibitem{parise2006strategies}
S.~Parise, R.~Cross, and T.~H. Davenport, ``Strategies for preventing a
  knowledge-loss crisis,'' \emph{MIT Sloan Management Review}, vol.~47, no.~4,
  p.~31, 2006.

\bibitem{kelly2015scientific}
D.~Kelly, ``Scientific software development viewed as knowledge acquisition:
  Towards understanding the development of risk-averse scientific software,''
  \emph{Journal of Systems and Software}, vol. 109, pp. 50--61, 2015.

\bibitem{latoza2006maintaining}
T.~D. LaToza, G.~Venolia, and R.~DeLine, ``Maintaining mental models: a study
  of developer work habits,'' in \emph{Proceedings of the 28th international
  conference on Software engineering}.\hskip 1em plus 0.5em minus 0.4em\relax
  ACM, 2006, pp. 492--501.

\bibitem{dingsoyr2014towards}
T.~Dings{\o}yr and N.~B. Moe, ``Towards principles of large-scale agile
  development,'' in \emph{International Conference on Agile Software
  Development}.\hskip 1em plus 0.5em minus 0.4em\relax Springer, 2014, pp.
  1--8.

\bibitem{rolland2016problematizing}
K.~H. Rolland, B.~Fitzgerald, T.~Dingsoyr, and K.-J. Stol, ``Problematizing
  agile in the large: alternative assumptions for large-scale agile
  development,'' in \emph{37th International Conference on Information
  Systems}, 2016.

\bibitem{sletholt2012we}
M.~T. Sletholt, J.~E. Hannay, D.~Pfahl, and H.~P. Langtangen, ``What do we know
  about scientific software development's agile practices?'' \emph{Computing in
  Science \& Engineering}, vol.~14, no.~2, pp. 24--37, 2012.

\bibitem{santos2015fostering}
V.~Santos, A.~Goldman, and C.~R. De~Souza, ``Fostering effective inter-team
  knowledge sharing in agile software development,'' \emph{Empirical Software
  Engineering}, vol.~20, no.~4, pp. 1006--1051, 2015.

\bibitem{santos2012knowledge}
V.~R. Santos, A.~L. Soares, and J.~{\'A}. Carvalho, ``Knowledge sharing
  barriers in complex research and development projects: an exploratory study
  on the perceptions of project managers,'' \emph{Knowledge and Process
  Management}, vol.~19, no.~1, pp. 27--38, 2012.

\bibitem{mesh2013scientific}
E.~S. Mesh and J.~S. Hawker, ``Scientific software process improvement
  decisions: A proposed research strategy,'' in \emph{Software Engineering for
  Computational Science and Engineering (SE-CSE), 2013 5th International
  Workshop on}.\hskip 1em plus 0.5em minus 0.4em\relax IEEE, 2013, pp. 32--39.

\bibitem{heaton2015claims}
D.~Heaton and J.~C. Carver, ``Claims about the use of software engineering
  practices in science: A systematic literature review,'' \emph{Information and
  Software Technology}, vol.~67, pp. 207--219, 2015.

\bibitem{segal2008scientists}
J.~Segal, ``Scientists and software engineers: A tale of two cultures,'' 2008.

\bibitem{kelly2007software}
D.~F. Kelly, ``A software chasm: Software engineering and scientific
  computing,'' \emph{IEEE Software}, vol.~24, no.~6, pp. 120--119, 2007.

\bibitem{hara2003emerging}
N.~Hara, P.~Solomon, S.-L. Kim, and D.~H. Sonnenwald, ``An emerging view of
  scientific collaboration: Scientists' perspectives on collaboration and
  factors that impact collaboration,'' \emph{Journal of the Association for
  Information Science and Technology}, vol.~54, no.~10, pp. 952--965, 2003.

\bibitem{jakobsen2004barriers}
C.~H. Jakobsen, T.~Hels, and W.~J. McLaughlin, ``Barriers and facilitators to
  integration among scientists in transdisciplinary landscape analyses: a
  cross-country comparison,'' \emph{Forest Policy and Economics}, vol.~6,
  no.~1, pp. 15--31, 2004.

\bibitem{hinsen2015technical}
K.~Hinsen, ``Technical debt in computational science,'' \emph{Computing in
  Science \& Engineering}, vol.~17, no.~6, pp. 103--107, 2015.

\bibitem{anderson2007perverse}
M.~S. Anderson, E.~A. Ronning, R.~De~Vries, and B.~C. Martinson, ``The perverse
  effects of competition on scientists’ work and relationships,''
  \emph{Science and engineering ethics}, vol.~13, no.~4, pp. 437--461, 2007.

\bibitem{van2012intended}
H.~P. Van~Dalen and K.~Henkens, ``Intended and unintended consequences of a
  publish-or-perish culture: A worldwide survey,'' \emph{Journal of the
  Association for Information Science and Technology}, vol.~63, no.~7, pp.
  1282--1293, 2012.

\bibitem{ratcheva2009integrating}
V.~Ratcheva, ``Integrating diverse knowledge through boundary spanning
  processes--the case of multidisciplinary project teams,'' \emph{International
  Journal of Project Management}, vol.~27, no.~3, pp. 206--215, 2009.

\bibitem{post2000comparison}
D.~Post and L.~Cook, ``A comparison of software engineering practices used by
  the llnl nuclear applications codes and by the software industry,'' in
  \emph{Nuclear Explosive Code Developers Conference. Oakland, CA, Lawrence
  Livermore National Laboratory}, vol.~18, 2000.

\bibitem{cronbach1951coefficient}
L.~J. Cronbach, ``Coefficient alpha and the internal structure of tests,''
  \emph{psychometrika}, vol.~16, no.~3, pp. 297--334, 1951.

\bibitem{suri2011counting}
S.~Suri and S.~Vassilvitskii, ``Counting triangles and the curse of the last
  reducer,'' in \emph{Proceedings of the 20th international conference on World
  wide web}.\hskip 1em plus 0.5em minus 0.4em\relax ACM, 2011, pp. 607--614.

\bibitem{nerkar2005evolution}
A.~Nerkar and S.~Paruchuri, ``Evolution of r\&d capabilities: The role of
  knowledge networks within a firm,'' \emph{Management science}, vol.~51,
  no.~5, pp. 771--785, 2005.

\bibitem{boden2009bridging}
A.~Boden and G.~Avram, ``Bridging knowledge distribution-the role of knowledge
  brokers in distributed software development teams,'' in \emph{Proceedings of
  the 2009 ICSE Workshop on Cooperative and Human Aspects on Software
  Engineering}.\hskip 1em plus 0.5em minus 0.4em\relax IEEE Computer Society,
  2009, pp. 8--11.

\bibitem{kabo2014proximity}
F.~W. Kabo, N.~Cotton-Nessler, Y.~Hwang, M.~C. Levenstein, and J.~Owen-Smith,
  ``Proximity effects on the dynamics and outcomes of scientific
  collaborations,'' \emph{Research Policy}, vol.~43, no.~9, pp. 1469--1485,
  2014.

\bibitem{schossau2014sustainable}
J.~Schossau and G.~Wilson, ``Which sustainable software practices do scientists
  find most useful?'' in \emph{Proceedings of the 2nd Workshop on Sustainable
  Software for Science: Practice and Experiences (WSSSPE2)}, 2014.

\bibitem{carver2013self}
J.~Carver, D.~Heaton, L.~Hochstein, and R.~Bartlett, ``Self-perceptions about
  software engineering: A survey of scientists and engineers,'' \emph{Computing
  in Science \& Engineering}, vol.~15, no.~1, pp. 7--11, 2013.

\bibitem{pinto2018scientists}
G.~Pinto, I.~Wiese, and L.~F. Dias, ``How do scientists develop scientific
  software? an external replication,'' in \emph{2018 IEEE 25th International
  Conference on Software Analysis, Evolution and Reengineering (SANER)}.\hskip
  1em plus 0.5em minus 0.4em\relax IEEE, 2018, pp. 582--591.

\bibitem{szymczak2016position}
D.~Szymczak, S.~Smith, and J.~Carette, ``Position paper: A knowledge-based
  approach to scientific software development,'' in \emph{Software Engineering
  for Science (SE4Science), IEEE/ACM International Workshop on}.\hskip 1em plus
  0.5em minus 0.4em\relax IEEE, 2016, pp. 23--26.

\bibitem{smith2016advantages}
S.~Smith, T.~Jegatheesan, and D.~Kelly, ``Advantages, disadvantages and
  misunderstandings about document driven design for scientific software,'' in
  \emph{Software Engineering for High Performance Computing in Computational
  Science and Engineering (SE-HPCCSE), 2016 Fourth International Workshop
  on}.\hskip 1em plus 0.5em minus 0.4em\relax IEEE, 2016, pp. 41--48.

\bibitem{gil2014collaborative}
Y.~Gil, E.~Moon, and J.~Howison, ``No science software is an island:
  Collaborative software development needs in geosciences,'' in
  \emph{Proceedings of the 2nd Workshop on Sustainable Software for Science:
  Practice and Experiences (WSSSPE2)}, 2014.

\end{thebibliography}

\end{document}